\documentclass[conference]{IEEEtran}
\IEEEoverridecommandlockouts
\usepackage[utf8]{inputenc}
\usepackage{subfig}
\usepackage[ruled,linesnumbered]{algorithm2e}
\usepackage{dirtytalk}
\usepackage{listings}
\usepackage{float}
\usepackage{multirow}
\usepackage{makecell}
\usepackage{hyperref}

\usepackage{cite}
\usepackage{amsmath,amssymb,amsfonts}
\usepackage{algorithmic}
\usepackage{graphicx}
\usepackage{textcomp}
\usepackage{xcolor}
\def\BibTeX{{\rm B\kern-.05em{\sc i\kern-.025em b}\kern-.08em
    T\kern-.1667em\lower.7ex\hbox{E}\kern-.125emX}}
\begin{document}

\title{Impact of Combining Syntactic and Semantic Similarities on Patch Prioritization while using the Insertion Mutation Operators
}






\author{\IEEEauthorblockN{\{Mohammad Raihan Ullah, Nazia Sultana Chowdhury, Fazle Mohammed Tawsif\}}
\IEEEauthorblockA{\textit{Institute of Information and Communication Technology} \\
\textit{Shahjalal University of Science and Technology}\\
Sylhet, Bangladesh \\
}
}

\maketitle

\begin{abstract}
Patch prioritization ranks candidate patches based on their likelihood of being correct. The fixing ingredients that are more likely to be the fix for a bug, share a high contextual similarity. A recent study shows that combining both syntactic and semantic similarity for capturing the contextual similarity, can do better in prioritizing patches. 
In this study, we evaluate the impact of combining the syntactic and semantic features on patch prioritization using the Insertion mutation operators. This study inspects the result of different combinations of syntactic and semantic features on patch prioritization. As a pilot study, the approach uses genealogical similarity to measure the semantic similarity and normalized longest common subsequence, normalized edit distance, cosine similarity, and Jaccard similarity index to capture the syntactic similarity. It also considers Anti-Pattern to filter out the incorrect plausible patches. The combination of both syntactic and semantic similarity can reduce the search space to a great extent. 
Also, the approach generates fixes for the bugs before the incorrect plausible one. 
We evaluate the techniques on the IntroClassJava benchmark using Insertion mutation operators and successfully generate fixes for 6 bugs before the incorrect plausible one. So, considering the previous study, the approach of combining syntactic and semantic similarity can able to solve a total number of 25 bugs from the benchmark, and to the best of our knowledge, it is the highest number of bug solved than any other approach. The correctness of the generated fixes are further checked using the publicly available results of CapGen and thus for the generated fixes, the approach achieves a precision of 100\%.
\end{abstract}


\section{Introduction}
Automated Program Repair is the automatic repair of software bugs without the intervention of a human programmer. Automatic debugging and repairing of the software defects can make a programmer’s life a lot easier.
\let\thefootnote\relax\footnotetext{DOI reference number: 10.18293/SEKE2022-047}

Manual Debugging is a very difficult, time-consuming, and painful task. Study shows that the global cost of debugging is 312 billion dollars annually and software developers spend 50 percent of their time fixing bugs\cite{ref1}. 

Automated Program Repair (APR) reduces the bug-fixing effort by automatically generating patches that satisfy a specification e.g. Test cases\cite{ref3}. A patch is a modification applied to a program for fixing a bug. A typical APR approach takes a faulty program and a set of test cases as input (where the program fails at least one test case) and produces a patch to repair the fault (the patched program at least passes all the tests). This type of technique is called the generate-and-validate technique. Patches that are applied to a program could exist in the faulty program itself as well as in a non-local program. The study founds that up to 69\% of the repaired code fragments (in the form of code lines) can be obtained from both the local program and non-local programs \cite{ref5}, \cite{ref6}.

Throughout the years, several Automated Program Repair (APR) techniques have been introduced. Such as GenProg, HDRepair, CapGen, ssFix, SimFix, Elixir, etc \cite{3},\cite{4},\cite{5},\cite{6},\cite{7},\cite{8},\cite{9},\cite{10}. Many of these techniques e.g. SimFix, ssFix, CapGen, Elixir, etc \cite{5},\cite{6},\cite{8},\cite{12},\cite{13} are known as search-based APR. So, APR is often described as a search problem in a patch candidate space. In a candidate space, there are the fixing code fragments or patches that can be a possible fix for that particular bug. Testing the fixing codes earlier that are more likely to be the fix for the bug can reduce bug fixing time. Sorting candidate patches based on their probability of being the correct fix for the bug, is called patch prioritization.

The faulty program and the code fragment where the fix resides share a high contextual similarity. To capture these similarity, some method \cite{elixir} \cite{simFix} \cite{ssFix} uses Syntactic features and some method \cite{capGen} uses Semantic features. A recent study \cite{ref7} shows that combining Syntactic and Semantic features in patch prioritization can reduce bug fixing time and achieve higher precision. So in this study, we use the combination of Syntactic and Semantic features in Patch Prioritization and observe the impacts while working with \emph{Insertion} as the mutation operation. We also evaluate some different ways of combining those feature scores to calculate the final score and analyze their impact on Patch Prioritization. We evaluate the techniques on IntroClassjava benchmark \cite{introclassjava} which contains 297 real-life bugs. We were successfully able to solve six bugs with a precision of 100\%.

\section{Related Work}
Testing the candidate patches early which, are more likely to be the fix can help to reduce
the bug fixing time and give some treatment to the search space explosion problem
of search-based Automated Program Repair.

The most representative of the search-based APR category is GenProg \cite{genProg}, which searches for
correct patches via genetic programming algorithm. GenProg
constrains the genetic operations of mutation and crossover to operate only on the region of the program that is relevant to the error.  GenProg repairs bugs with the help of the following components: Mutation Space, Crossover, Evaluation Function. GenProg presents a program as an Abstract Syntax Tree (AST) and modifies
the faulty node using the mutation operations such as insertion, deletion, and Replacement operators. AST of the program is modified using mutation from
the mutation space. As a result of the modification, variants of the faulty program are
generated then they are passed to the evaluation function to see if those variants pass
the test cases.

Elixir\cite{elixir} generates patches based on generate-and-validate techniques. For a given bug,
Elixir takes as input a buggy program, a test suite, and optionally a bug report, then produces
a patch that passes all the test cases. Elixir works in the following ways: Bug Localization,  Generating Candidate Patches, Ranking and Selection of Candidate Patches, Validating Selected Candidate Patches. Elixir uses four syntactical features to
rank the patches. They are Distance score between faulty and fixing ingredients, Contextual Similarity, Frequency in the context, Bug Report Similarity. The approach introduces 8 templates and validates only the top 50 patches generated from
each template.

SimFix\cite{simFix} defines a method to obtain a search space from existing patches based on an
abstract space definition on AST types. The approach in SimFix consists of - an offline mining stage, an online repairing stage. SimFix uses the Ochiai algorithm to localize fault which produces an
ordered list of suspicious statements. It extracts a set of fixing ingredients
using syntactic features, such as Structure Similarity, Variable Name Similarity, Method Name Similarity. Then it matches variables in the faulty and fixes ingredients and builds a variable mapping table for code adaptation. It calculates the difference
between faulty and fixing ingredients and extracts concrete modification. Finally, it generates a list of patches and
validates them using the test suite. These three features obtain low precisions 63.41\%, 33.33\%, and 60.7\%, respectively.

ssFix\cite{ssFix} uses the syntactic features to calculate the similarity between faulty code and fixing
ingredients. It uses the TF-IDF model to calculate the similarity score. They evaluated their technique on 357 bugs in the Defect4j\cite{defect4j} bug dataset and successfully
solved 20 bugs from that dataset.

CapGen\cite{capGen} uses semantic features to prioritize patches. It works on the fine-grained level to
extract fixing ingredients. They selected 30 augmented mutation
operators to generate operation space. For example: insert Simple Name under Method
Invocation or delete Expression Statement under Method Declaration etc. CapGen uses three models to calculate the similarity score between faulty nodes and fixing
ingredients. They are Genealogy Context, Accessed Variable, Semantic Dependency. CapGen combines the three model scores to calculate the final score. Based
on that score, it prioritizes the patches and tests the patches which are more likely to be the
fix for the bug. This approach achieves a precision of 84\%.

A study [17] showed that combining syntactic and
semantic features for patch prioritization can work better than using syntactic and semantic
features alone. In the study, they took expression level bug and used replacement mutation
operation for patch generation. To
calculate the similarity between faulty code and fixing ingredients, it used Genealogical \& variable similarity as semantic metrics and Normalized Longest Common Subsequence (Normalized LCS) as syntactic metrics. The approach was evaluated using
22 bugs from the IntroClassJava benchmark\cite{introclassjava}, fixed by
CapGen applying replacement mutation\cite{capGen}. It repaired all 22 bugs and achieved a precision of 100\%.

\section{Methodology}
\begin{figure}[h]
    \subfloat[A fixed bug (Math 70 from Defects4J) using \emph{Insertion} mutation operation]{
    \label{code_1}\includegraphics[scale=0.3]{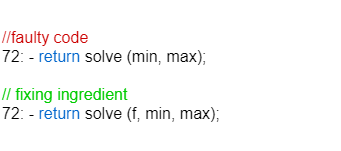}
    } \\
    \subfloat[AST difference of the fixed bug \ref{code_1}]{
    \label{code_1_tree}\includegraphics[scale=0.3]{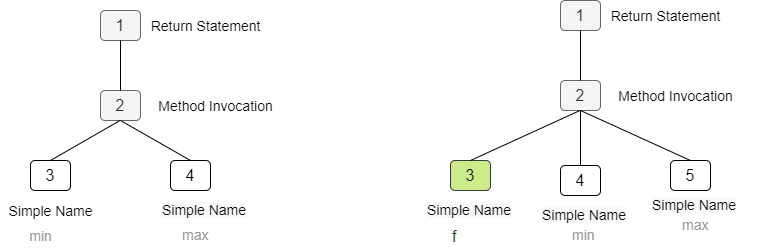}
    }
    \\
    \subfloat[A Fixed Bug (Lang 43 in Defects4J) Using the \emph{Insertion} Mutation Operation and Fix Extracted From the Same File]{
    \label{code_2}\includegraphics[scale=0.3]{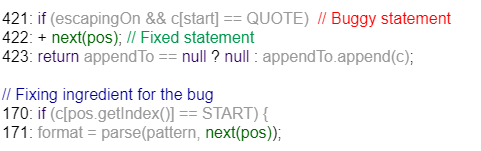}
    }
    \caption{Sample Bug Fixes using Insertion Mutation Operation}
    \label{insertion-example-1}
\end{figure}

A recent study \cite{ref7} shows that combining both syntactic and semantic similarity can improve patch prioritization. It works at \emph{Expression} level and uses \emph{Replacement} mutation operators to generate the patches. In this research, we evaluate the technique further to observe its impact on patch prioritization while using the \emph{Insertion} as the mutation operation. The approach works at the \emph{Expression} and \emph{Expression-Statement} level as the \emph{Insertion} mutation operation works better at these level \cite{capGen}. It manipulates the source codes representing them in Abstract Syntax Tree (AST). Figure \ref{insertion-example-1} shows some sample bug fixes using \emph{Insertion} Mutation Operation. In figure \ref{code_1}, the bug is fixed by inserting a \emph{Simple Name} under the faulty \emph{Method Invocation}. In figure \ref{code_2}, the bug is fixed inserting a \emph{Method Invocation} under the \emph{If-Statement}. 

The approach in this paper takes a source program and a test suite (positive and negative test cases) as input and generates some repaired variants. For generating patches, it uses \emph{Insertion} mutation operation using \emph{Expression} and \emph{Expression-Statement} level code for fixing elements. It uses Cosine Similarity, Jaccard Similarity Index, Normalized Edit Distance, Normalized LCS as a syntactic feature, and Genealogical similarity and Anti-Pattern as a semantic feature. The repaired variants are thus tested with the test suite and the correct patches for the bug are validated.
The approach works in the four following steps. The steps are quite similar to the approach of the previous study \cite{ref7}.

\subsection{Fault Localization}
This step identifies the faulty code using Ochiai \cite{ochai} algorithm, a spectrum-based fault localization technique. It takes help from the GZoltar \cite{GZoltar} tool for this purpose. GZoltar takes a source code and a test suite (including positive and negative test cases) and generates a list of the suspicious statements.
For \emph{Insertion} mutation operation, the approach identifies the faulty AST node of \emph{Expression} and \emph{Expression Statement}. It outputs a line number-wise suspicious value (probability of being a faulty node) of a program. If a node spans across multiple lines, it is assigned the average of the suspicious values of those lines.

\subsection{Patch Generation}
In this step, the approach generates patches inserting fixing ingredients under the faulty node. For faulty nodes, it takes \emph{Expression} and \emph{Expression-Statement} level bug. Ambiguity arises while inserting a node under a block statement. This ambiguity can be addressed as \emph{fix location selection}  problem. To avoid ambiguity for fix location, it further splits the insertion operation into two more categories: 
\begin{itemize}
    \item Insert Before: Insert a subtree before node \emph{N}. 
    \item Insert After : Insert a subtree after node \emph{N}. 
\end{itemize}

The operation space induced by \emph{Insertion} mutation operator is huge. To reduce this space, we use the top 10 augmented insertion mutation operations from \emph{CapGen} \cite{capGen}.

\subsection{Patch Prioritization}
\label{prio}
Several patches are produced at the patch generation level. To minimize the bug fixing time and cost, patches are prioritized. It also helps to maximize the precision of patches. For
prioritizing patches, the genealogical similarity between is used to measure semantic similarity, and
normalized Longest Common Subsequence (LCS) and Anti - Pattern \cite{ref8} and Cosine Similarity is used to measure
syntactic similarity.
\begin{itemize}
    \item \textbf{Genealogical Similarity: } The genealogical structure of an AST node shows a node is frequently used under and together with which types of code elements by checking its ancestor \& sibling node \cite{capGen}. Traversal is done from the node to its ancestor until a method declaration is found. And nodes of category Statement and Expression within the same Block are extracted for sibling nodes\cite{capGen}. The Genealogical Similarity is measured as 
  \begin{equation}
 gen\_s(f_n, f_e) = \dfrac{\sum_{t\epsilon k} \min{(f_n(t),f_e(t))}} {\sum_{t\epsilon k} f_n(t)}  
  \end{equation}\\
    Here, \(f_n\) and \(f_e\) are frequencies of different node types for faulty node and fixing ingredient respectively. K indicates a set of all distinct AST node types captured by \(f_n\)

    \item \textbf{Normalized LCS:} LCS measures the Longest Common Subsequences between two sets of sequence. To measure the LCS, we take string representation of faulty node \( (f_n) \) and fixing ingredient\( (f_e) \). Then the value of the LCS is normalized to get the value in the interval [0,1]. 
  \begin{equation}
      n\_lcs((f_n, f_e)) = \dfrac{LCS(f_n,f_e)}{\max{(f_n,f_e)}} 
  \end{equation}
  
    \item \textbf{Cosine Similarity:} The cosine similarity of two ASTs representative vectors expresses the similarity between them.
The cosine similarity measure uses two finite-dimensional vectors   of   the   same   dimension   in   which   each   vector represents a document. Given two string \( f_n \) (string representation of faulty node) and \( f_e \) (string representation of fixing ingredient), we construct the collection between the two elements. The collection   of   terms   denotes C =   \{\(c_1\),   \(c_2\),...,  \(c_n\)\}   with \(c_{i} \in f_n \mid c_{i} \in f_e \) and each \(c_{i}\) is distinct. The strings are then represented in an n-dimensional vectors \( \vec{V}_{\,f_n} \) and \( \vec{V}_{\,f_e} \).

\begin{equation}
 cos\_s(f_n, f_e) = \dfrac{\vec{V}_{\,f_n} . \vec{V}_{\,f_e}} {(|\vec{V}_{\,f_n}| * |\vec{V}_{\,f_e}|)}
  \end{equation}
  
    \item \textbf{Normalized Edit Distance}: Edit distance refers to the minimum number of edit operations needed to convert from one string to another string. For measuring the score, we use an alternative version of Levenshtein distance where each Insertion or Deletion cost is 1 and Substitution cost is 2. The score is then normalized to get the score between the interval [0,1]. Finally, the score is subtracted from 1 as the approach in this paper only considers the similarities between the faulty nodes and fixing ingredients. 
    
     \begin{equation}
      n\_ed(f_n, f_e) = 1 - \dfrac{Edit\_Distance(f_n,f_e)}{\max{(f_n,f_e)}} 
  \end{equation}\\
  
  Here, \( f_n \) and \( f_e \) is the string representation of faulty node and fixing ingredient.
  
    \item \textbf{Jaccard Similarity Index:} The technique computes Jaccard distance between string representation of faulty node \( (f_n) \) and fixing node \( (f_e) \). The input string is converted into a set of N-grams. In Comparing with Cosine similarity, Jaccard Similarity Index takes the number of common attributes is divided by the number of attributes that exist in at least one of the two objects. In comparison, Cosine similarity divides the number of the common attributes by the dot product of two objects represented in vectors. 
\begin{equation}
jac\_s(f_n, f_e) = \dfrac{|f_n \cap f_e| }{|f_n \cup f_e|}
\end{equation}\\
Here, \(f_n\) and \(f_e\) is the string representation of faulty node and fixing ingredient.

    \item \textbf{Anti-Pattern:} Anti-patterns capture disallowed
modifications to the buggy program \cite{antiPattern}. If
such a modification passes all tests in the given test-suite, they are not counted as repairs \cite{antiPattern}. Because those modifications may later introduce some code smell in the program. We filter out such modifications as part of patch prioritization. We take account of the following anti-pattern \cite{antiPattern} for Insertion mutation operation. 

\begin{itemize}
    \item \emph{Anti-append Early Exit}: This pattern disallows the insertion of return statements at any location except for after the last statement.
    
\end{itemize}
The utility of anti-pattern is not directly used with the semantic-based counterparts. Rather it is used to filter out the trivial solutions. 
\end{itemize}

After calculating the syntactic and semantic features, they are then combined to calculate the final score. We keep Genealogical Similarity and Normalized LCS fixed and used Cosine Similarity, Normalized Edit Distance, Jaccard Similarity Index one at a time. So, we evaluated total 3 combinations of features. The first one was \textbf{Com-CS}, the combination of Genealogical Similarity, Normalized LCS and Cosine Similarity. The second one was \textbf{Com-NED}, the combination of Genealogical Similarity, Normalized LCS and Normalized Edit Distance. And the final one was \textbf{Com-JS}, the combination of Genealogical Similarity, Normalized LCS and Jaccard Similarity Index.

We calculate the score of the patches in the following way:\\
\begin{equation}
    patch\_score = gen\_s(f_n,f_e) + n\_lcs(f_n,f_e) + simi(f_n,f_e) 
\end{equation}\\
Here, \emph{simi\( (f_e, f_n) \)} stands for Cosine Similarity, Normalized Edit Distance, Jaccard Similarity Index, and Jaccard Similarity Index which we plugged into the equation one at a time. 

\subsection{Patch Validation}
Patch validation validates the patch that is the correct repair of the faulty node. This step executes negative and positive test cases for validating the patches.

\section{Experiment}
We implemented the technique addressed in this paper in Java. It uses the Eclipse JDT parser for manipulating AST. It uses the GZoltar\cite{GZoltar} tool to localize the faulty codes. GZoltar tool is used with Ochiai algorithm \cite{ochai} to calculate the suspicious value of the \emph{Expression} and \emph{Expression-Statement} for being a faulty node. The pseudocode of our technique is given in \ref{algo}. 

\begin{algorithm}[h]
\SetAlgoLined
\SetKwInOut{Input}{Input}
\SetKwInOut{Output}{Output}
\Input{Program \emph{P} to be repaired; Set of positive test cases, \emph{PosT}; Set of negative test cases, \emph{NegT}; Set of augmented mutation operation, \emph{AugOp.}}
\Output{Repaired program variant}
$RepairedProgramVariants \gets \emptyset$ \\
$\phi \gets localize\_fault(P, PosT, NegT)$ \\
\ForAll{$faultyNode \in \phi$}{
$\theta \gets getFixingElement()$ \\
$N \gets getParent(faultyNode)$ \\
\ForAll{$fixingIngredient \in \theta $}{
$OP \gets getMutationOperation(N, fixingIngredient)$
}
}
\ForAll{$fixingIngredient \in \theta \And  OP \in AugOp $}{
$syntacticSimi \gets calculateSyntacticSimi(faultyNode, fixingIngredient, OP)$ \\
$semanticSimi \gets calculateSemanticSimi(faultyNode, fixingIngredient, OP)$ \\
$finalScore \gets combine(syntacticSimi, semanticSimi)$ \\
$candidatePatches \gets \{faultyNode, fixingIngredient, finalScore\}$\\
}
$sortedCandidatePatches \gets sort(candidatePatches)$ \\
\ForAll{$patch \in sortedCandidatePatches$}{
$programVariant \gets insert(faultyNode, fixingIngredient, AugOp)$ \\
$RepairedProgramVariants \gets validate(programVariant, PosT, NegT)$
}
\Return RepairedProgramVariants
\caption{High-level pseudocode for our technique.}
\label{algo}
\end{algorithm}
\subsection{Research Question}
As we discussed in the earlier section, the existing patch prioritization technique can be categorized into 3 broad categories. Recent work\cite{ref7} introduced the technique for combining syntactic and semantic similarity for patch prioritization, worked at the \emph{Expression} level bug and used the \emph{Replacement} mutation operation to generate patches. In this study, we evaluate the technique of combining syntactic and semantic similarity to observe its impact on patch prioritization using the \emph{Insertion} mutation operation. Our evaluation aims to answer the following research questions:
\begin{itemize}
    \item \textbf{RQ1: }What is the impact of combining syntactic and semantic similarity while using \emph{Insertion} mutation operation?
    
    
    \item \textbf{RQ2: }How different syntactic and semantic similarity features impact patch prioritization?
\end{itemize}\subsection{Evaluation}
We evaluated our technique in IntroClassJava Benchmark \cite{introclassjava}. The benchmark contains 297 bugs from 6 projects. For evaluation, the following are examined:
\begin{itemize}
    \item \textbf{Median Rank of the Correct Patch:} The lower
the median rank, the better the approach is \cite{6}.
    
    \item \textbf{Precision:} If a technique can rank the correct solution before the incorrect plausible one, it achieves higher precision. \cite{capGen}
    
    \item \textbf{Rank of the first correct solution:} The approach that
ranks the first correct solution higher, considered as more efficient. \cite{6}.



\end{itemize}
\subsection{Experiment Settings}
To understand the impact of similarities, the approach need to compared with patch prioritization techniques that use semantic and syntactic similarity individually. Therefore, this study
further implements semantic or syntantic similarity
based patch prioritization approaches using metrics
discussed in Section \ref{prio}. The techniques are described below: 
\begin{itemize}
    \item \textbf{Semantic Similarity based Approach (SSBA):} It uses only semantic similarity metrics namely
genealogical to prioritize
patches. Also the anti-pattern is also used to filter out the incorrect patches.

\item \textbf{LCS based Approach (LBA):} It is a syntactic
similarity based approach that prioritizes patches
using only normalized LCS score.

\item \textbf{Cosine Similarity based Approach (CSBA):} It is another syntactic
similarity based approach that prioritizes patches
using only normalized Cosine score.

\item \textbf{Jaccard Similarity Index based Approach (JSBA):} It is also a syntactic similarity based approach that uses only Jaccard Similarity Index score

\item \textbf{Normalized Edit Distance based Approach (NBA): }It is also another syntactic
similarity based approach that prioritizes patches
using only normalized Edit Distance score.
\end{itemize}
All of these five patch prioritization approaches
follow the same repairing process as the combined
ones. It ensures that the observed effects occurred
due to varied similarities used in patch prioritization.

\begin{table*}[t]
\centering
\begin{tabular}{ |c|c|c|c|c|c|c| } 
 \hline
 \multirow{2}{*}{\textbf{Bug Id}} & \multicolumn{2}{|c|}{\textbf{With Cosine Similarity}} & \multicolumn{2}{|c|}{\textbf{With Normalized Edit Distance}} & \multicolumn{2}{|c|}{\textbf{With Jaccard Similarity Index}} \\
 \cline{2-7}
  & \textbf{Total Patches} & \textbf{Correct Patch Rank} & \textbf{Total Patches} & \textbf{Correct Patch Rank} & \textbf{Total Patches} & \textbf{Correct Patch Rank}\\
 \hline
digits\_07045530\_002.java & 57 & 27 & 78 & 51 & 57 & 34\\
\hline
median\_0cdfa335\_003.java & 137 & 44,68 & 170 & 60 & 137 & 59, 63\\
\hline
median\_89b1a701\_010.java & 145 & 30 & 185 & 27 & 145 & 35\\
\hline 
smallest\_6aaeaf2f\_001.java & 197 & 90 & 226 & 27 & 197 & 86\\
\hline 
smallest\_818f8cf4\_003.java & 199 & 46 & 178 & 46 & 199 & 48\\
\hline 
smallest\_cb243beb\_000.java & 168 & 96 & 189 & 90 & 168 & 93\\
\hline
\end{tabular}
\caption{Results of the Approach of Combining Features on IntroClassJava Benchmark}
\label{finalApproach}
\end{table*}

\begin{table*}[t]
\centering
\begin{tabular}{ |c|c|c|c|c|c|c|c|c|c|c| } 
 \hline
 \multirow{2}{*}{\textbf{Bug Id}} & \multicolumn{2}{|c|}{\textbf{SSBA}} & \multicolumn{2}{|c|}{\textbf{LBA}} & \multicolumn{2}{|c|}{\textbf{CSBA}} & \multicolumn{2}{|c|}{\textbf{JSBA}} & \multicolumn{2}{|c|}{\textbf{NBA}} \\
 \cline{2-11}
  & \textbf{TP} & \textbf{CPR} & \textbf{TP} & \textbf{CPR} & \textbf{TP} & \textbf{CPR} & \textbf{TP} & \textbf{CPR} & \textbf{TP} & \textbf{CPR}\\
 \hline
digits\_07045530\_002.java & 326 & 107 & 357 & 342 & 58 & 23 & 58 & 42 & 82 & 67\\
\hline
median\_0cdfa335\_003.java & 652 & 110, 120 & 682 & 287, 340 & 138 & 42, 92 & 138 & 52, 85 & 175 & 81\\
\hline
median\_89b1a701\_010.java & 660 & 299 & 645 & 26 & 146 & 26 & 146 & 37 & 189 & 26\\
\hline 
smallest\_6aaeaf2f\_001.java & 817 & 170 & 870 & 113 & 198 & 130 & 198 & 140, 151 & 233 & 26\\
\hline 
smallest\_818f8cf4\_003.java & 779 & 157, 175 & 851 & 48, 414 & 200 & 70 & 200 & 58 & 181 & 50\\
\hline 
smallest\_cb243beb\_000.java & 589 & 310 & 631 & 241 & 168 & 124 & 168 & 121 & 190 & 110\\
\hline
\end{tabular}
\caption{Results of the Approach using Only Semantic or Syntactic Features Alone on IntroClassJava Benchmark \newline(TP = Total Patches, CPR = Correct Patch Ranks)}
\label{table2}
\end{table*}

\subsection{Result Analysis}
The approach repairs 6 bugs from IntroClassJava benchmark \cite{introclassjava}. For the repaired bugs, no plausible patch is generated before the first correct
solution. Thus, the approach achieves a precision of 100\%.
The implementation is publicly available \href{https://github.com/CosmicBeing09/Impact-of-Syntactic-and-Semantic-Similarities-On-Patch-Prioritization-Using-Insertion-Operation}{here}.


In our study, we were able to repair 6 bugs from the IntroClassJava benchmark. The results are shown in the table \ref{finalApproach}

\begin{figure}[h]
    \centering
    \fbox{\includegraphics[scale=0.28]{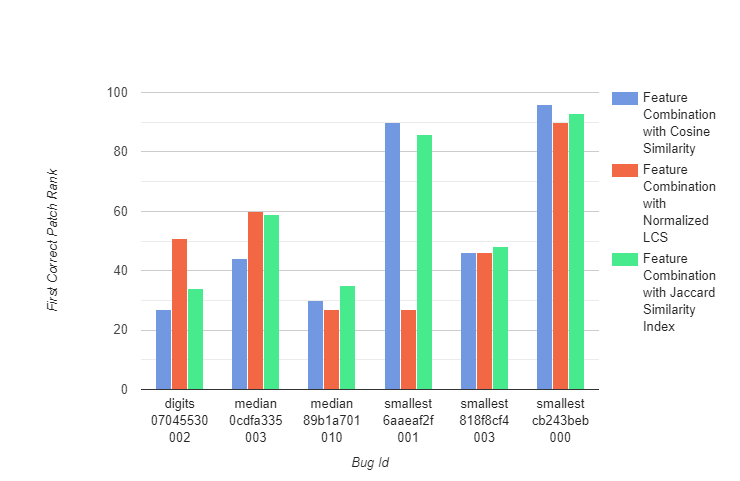}}
    \caption{Impact of different combination of features on patch rank}
    \label{fig:firstCorrectRank}
\end{figure}

\begin{figure}[h]
    \centering
    \fbox{\includegraphics[scale=0.28]{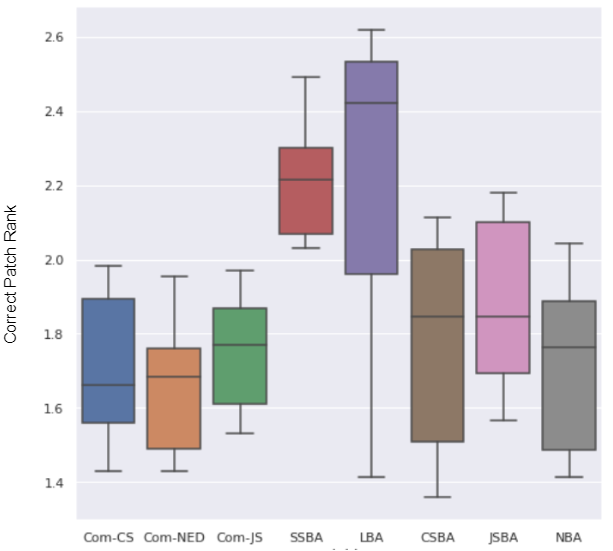}}
    \caption{Comparison of Correct Patch Rank among various Approach}
    \label{fig:boxplot}
\end{figure}

Figure \ref{fig:firstCorrectRank} shows the first correct patch rank of different approach. The lower the rank is, the better the approach.
Figure \ref{fig:boxplot} shows the rank distribution of Com-CS, Com-NED, Com-JS, SSBA, LBA, CSBA, JSBA, and NBA. Log transformation is used in this figure as the data range is high. 
In terms of median rank, Com-CS, Com-NED, and Com-JS outperform SSBA, LBA, CSBA, JSBA and NBA. The ranks are 45, 48.5, 53.5, 163.5, 177, 55, 55, 58.5 respectively. The reason is combination of features incorporate information from multiple
domains (both textual similarity and code meaning).
The correctness of the generated patches are checked using the publicly available results of CapGen \cite{capGen} except for the bug \emph{smallest\_cb243beb\_000}. CapGen doesn't contain the fix for that bug. The study \cite{ref8} that shows that combining both semantic and syntactic similarities works better than the syntactic or semantic similarities alone, was able to fix 22 bugs using Replacement Mutation Operation. In this paper we evaluate the impact if we use Insertion Mutation Operation. So, in total 25 bugs have been solved from the \emph{IntroClassJava} benchmark \cite{introclassjava} using the approach of combining syntactic and semantic similarities.  


\section{Conclusion}
In this paper, a patch prioritization technique is evaluated for combining syntactic and semantic similarity for automated program repair. Our approach uses different combinations of syntactic and semantic matrices to see their impact on patch prioritization while using the Insertion mutation operation for patch generation. It uses \emph{Genealogical Similarity}, \emph{Normalized LCS}, \emph{Cosine Similarity}, \emph{Jaccard Similarity Index}, \emph{Normalized Edit Distance} to calculate the context similarity score. It also considers \emph{Anti-Patterns} to filter out the incorrect plausible patches. The approach observes that \emph{Cosine Similarity} and \emph{Normalized Edit Distance} along with \emph{Genealogical Similarity} and \emph{Normalized LCS} rank the correct patch higher. This approach has been able to solve 6 bugs from IntroClassJava benchmark \cite{introclassjava}. It achieves a precision of 100\% for solving those bugs.

\end{document}